# How Students Use Media: A Comparison Across Faculties

Gerd Gidion[1], Luiz Fernando Capretz[2], Ken N. Meadows[2], and Michael Grosch[1]

[1]Karlsruhe Institute of Technology, Karlsruhe, Germany

[1]Western University, London, Ontario, Canada

*Abstract*—The pervasiveness of online information services has led to substantial changes in higher education including changes in faculty members' teaching methods and students' study habits. This article presents the results of a survey about media use for teaching and learning conducted at a large Canadian university and highlights trends in the use of new and traditional media across university Faculties. The results of this study support the assumption that student media usage includes a mixture of traditional and new media.

## I. INTRODUCTION

A Media Usage Survey was developed to provide researchers with a detailed understanding of student' technology use in learning and of possible environmental factors that may influence that use. This survey incorporated the entire spectrum of media services, and focused on creating a knowledge base for universities to understand the media usage of students and to establish a longitudinal international survey on technology use in tertiary education.

## II. RESEARCH METHODOLOGY

The survey uses an anonymous questionnaire format consisting of 150 items. Specifically, the tool measures usage frequency and user satisfaction with 53 media services, including information services such as Google, library catalogues, printed books, e-books, printed journals, e-journals, Wikipedia, open educational resources, and bibliographic software. In addition, the tool measures communication services, such as Twitter and Facebook; e-learning services, such as learning platforms and wikis; and media hardware such as tablets, desktop computers, smartphones and mobile learning [1]. Partial results of this survey showing students' satisfaction with media is described in [2], for engineering students only is detailed in [3], and a multicultural study is show in [4].

The survey tool was first developed in 2009 and used at the Karlsruhe Institute of Technology (KIT). During the course of 15 follow-up surveys that were administered on an international basis, the original survey underwent optimization, translation into several languages, and validation. In this study, the survey was administered at Western University in London, Ontario, with undergraduate students. The survey, which resembles a student questionnaire, intends to compare the media usage of students by examining possible divergences in media culture that may create problems in the use of media for studying and learning.

In the period between January 16th and February 15th 2013, a total of 19,978 students were invited to respond to the survey. Subsequently, exactly 1,584 visits occurred at the survey website. Among the invited students, 1266 started to answer the questions, 985 completed the survey, and 803 submitted, a completion rate of more than 90%.

Western researchers, in cooperation with research team members at the KIT, conducted the standardized and comparative analysis. Initial invitations to participate in the research and two reminders were sent by email. Both faculty and student surveys were voluntary and anonymous, as indicated in the letter that were distributed. For the student survey, three e-mails were sent by Office of the Registrar staff to a stratified random sample of undergraduate and graduate students enrolled on the main campus in the Winter 2013 academic term. The data for this survey was collected online using Unipark.

Table I. Response numbers for those who answered the question: Which faculty offers your primary area of study?

| Which faculty offers your primary area of study? | | Students | |
|---|---|---|---|
| | | Frequency | Percentage |
| **Respondents** | Science | 173 | 21.9 |
| | Engineering | 56 | 7.1 |
| | Health Science | 125 | 15.8 |
| | Social Science | 237 | 30.0 |
| | Arts&Humanities | 82 | 10.5 |
| | Other | 116 | 14.7 |
| | Sub-Total | 789 | 100.0 |
| | Missing | 196 | |
| | Total | 985 | |

### III. FINDINGS

Analysis of the survey requires interpretation of additional information concerning the current habits of media use for studying and teaching. In addition, the survey presents several limitations, including the reality that survey results fail to explain facts or observable behavior; rather, the answers merely analyze participants' responses to the questions.

#### A. General Media, Learning, and Studying Habits

The differentiation between students regarding their faculties shows just a few significant results, and these only between two certain groups (e.g. students from Engineering state that they study in groups more frequently than students from Arts & Humanities and Social Sciences, which might even be surprising). In general, the results indicate that students from all faculties attend class most frequently, followed by studying using a computer, and by themselves at home, as shown if Figure 1. All students use to search the internet for learning materials, and students from Engineering visit libraries less than other students.

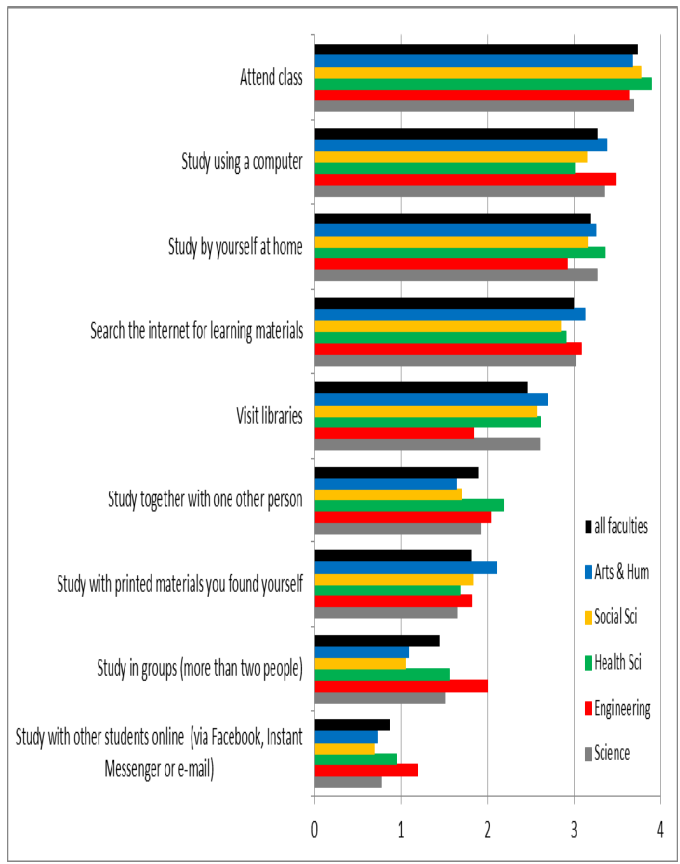

Fig. 1. Means of students' responses to the question: How often do you do the following? The question was rated on a five-point Likert scale with the following choices: "never" (0), "rarely" (1), "sometimes" (2), "often" (3), and "very often" (4); the figure shows the means of all students (valid n = 985) and of 5 subgroups, related to their faculty (Arts and Humanities valid n = 82, Engineering valid n = 56, Health Sciences valid n = 125, Science valid n = 173 and Social Science valid n = 237) (sorted by the means of all faculties).

#### B. Frequency of IT-Services Usage

The answers from students in different faculties were usually similar in the areas of usage frequency of various IT-devices for learning and studying. The one different is that Desktop-PCs and Computer Labs on Campus are often used by Engineering, as depicted in Figure 2.

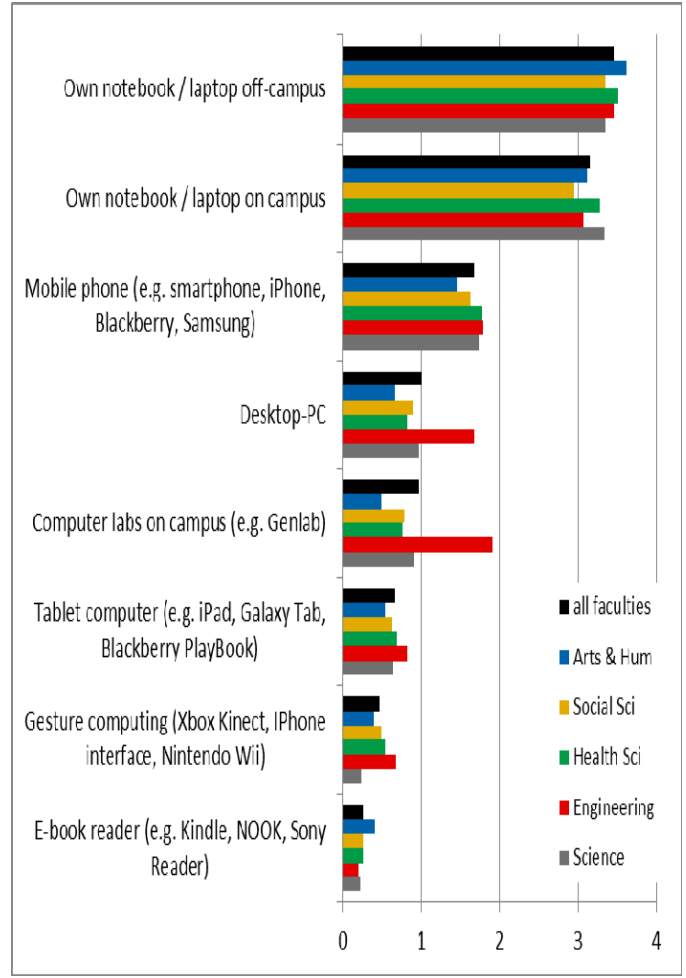

Fig. 2. Means of students' responses to the question: How often do you use the following for learning/studying?

#### C. Usage of Applications Related to Social Network

The survey results concerning the usage frequency of applications related to the social network is displayed together with the result for Google search, because this seems to be a necessary measure for the relevance of the other items. So, Facebook – the Internet application with a very high frequency in free time usage – is used far less frequently than Google search for study purposes, and Wikipedia seems to be even more relevant up to the present. Significant differences in the use of Wikipedia were evident, with the students from health sciences reporting less frequent use for learning than students from Science and Engineering. Students from Health Sciences

175

stated a slightly higher frequency of usage for the items related to Facebook (compared to students from Arts & Humanities) and Google+ (compared to students from Engineering and from Science). Please refer to Figure 3.

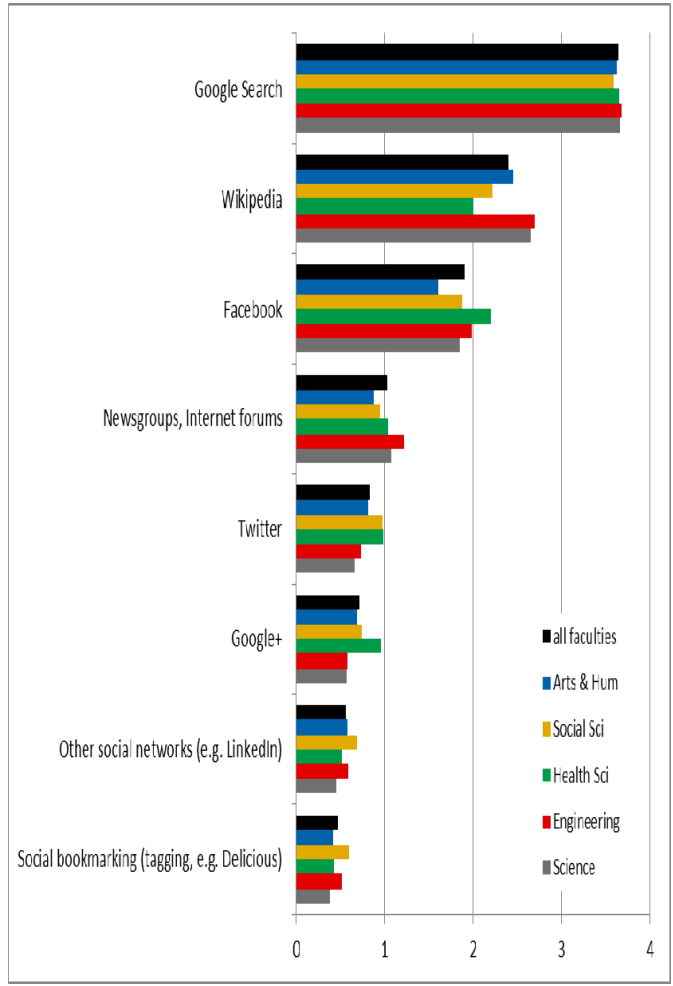

Fig. 3. Means of Students' responses to the question: How often do you use the following for learning/studying?

### D. Frequency of e-Learning Applications

The usage frequency of diverse e-Learning applications is headed by video sharing websites (but on a notedly lower level than in freetime use), which in general are utilized more than recorded lectures. Significant faculty differences are apparent for the item "online (self) tests for studying", with Arts and Humanities students being less frequent users than students from all of the other three faculties with which they were compared.

There also were significant differences on the use of recorded lectures, with Science students reporting more frequent usage than the other students, this may be because that faculty is reported to be very active in this field and offers several services and contents around recorded lectures. This is described in Figure 4.

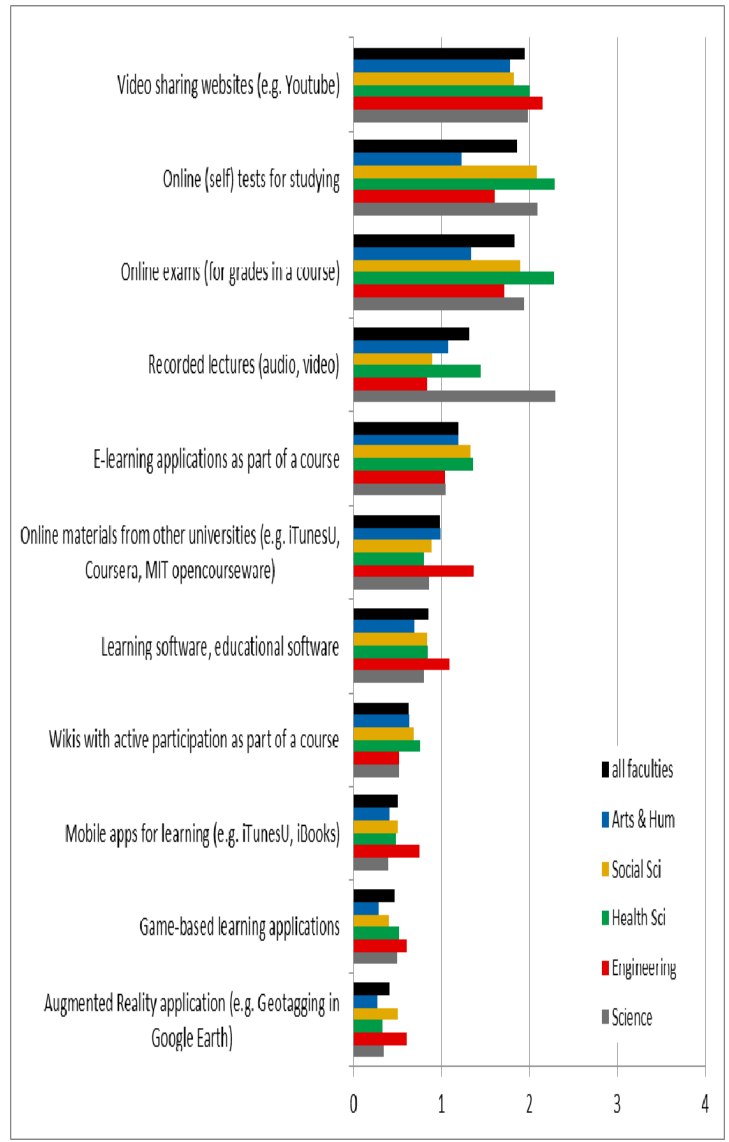

Fig. 4. Means of students' responses to the question: How often do you use the following for learning/studying?

### E. Frequency of Printed Versus Electronic Media Usage

The usage frequency of printed media compared with electronic media shows similar results for online material and / or scientific articles from instructors. Significant faculty differences were found for the items "online slides (from instructors)" with students from science and health science reporting a higher frequency of usage than students from arts and humanities and social science. Printed books, academic journals, online dictionaries and Google books seem to be more frequently used by students from Arts & Humanities compared to students from some other faculties, and students from Engineering show a higher mean of usage frequency for ebooks. These results are graphically displayed in Figure 5, in the next page.

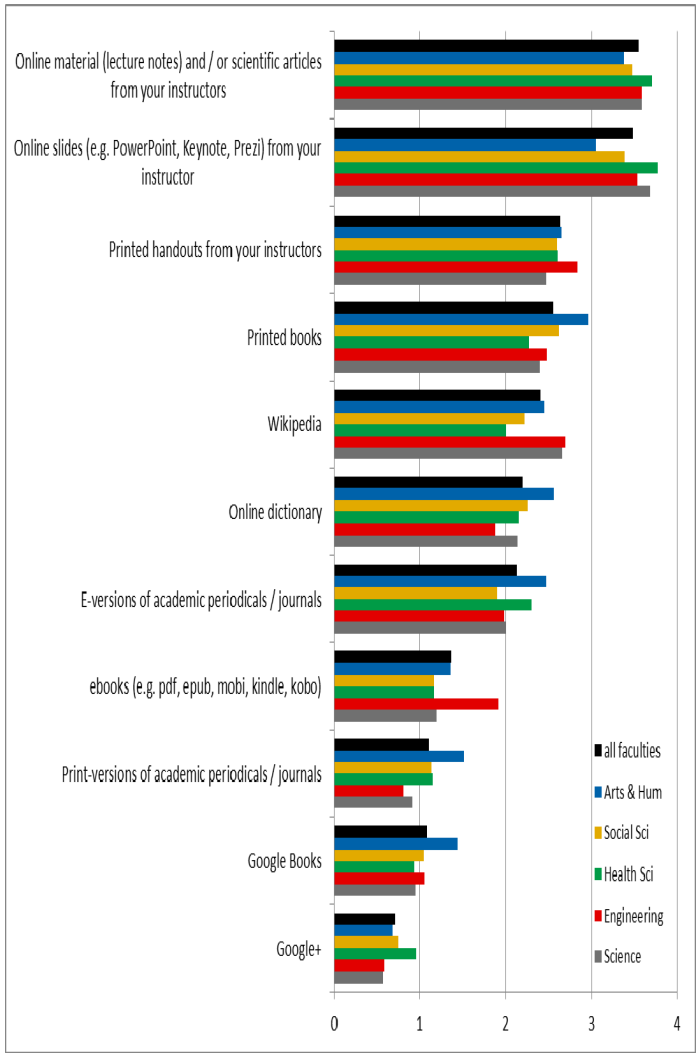

Fig. 5. Means of students' responses to the question: How often do you use the following for learning/studying?

*F. Media Usage in Free Time*

The students were asked about their use of diverse media in their free time. The results generally show an intensive use of Facebook and video-sharing websites (e.g., YouTube), as shown in Figure 6. Reading books and watching TV, two traditional media habits, were used only moderately. Certain media usage, like playing computer games, is, for most students, less relevant. Very new media, such as Google+, seems not to be relevant, at least at the time of the survey. To determine if there were differences between students from different faculties, we compared students from the five largest faculties.

We did find significant differences ($p < .01$) between the faculties on two items: for the item "Read books", Arts and Humanities students reported reading more books than all other students, and for the item "Play computer games", engineering students reported a higher frequency of playing games than health science students.

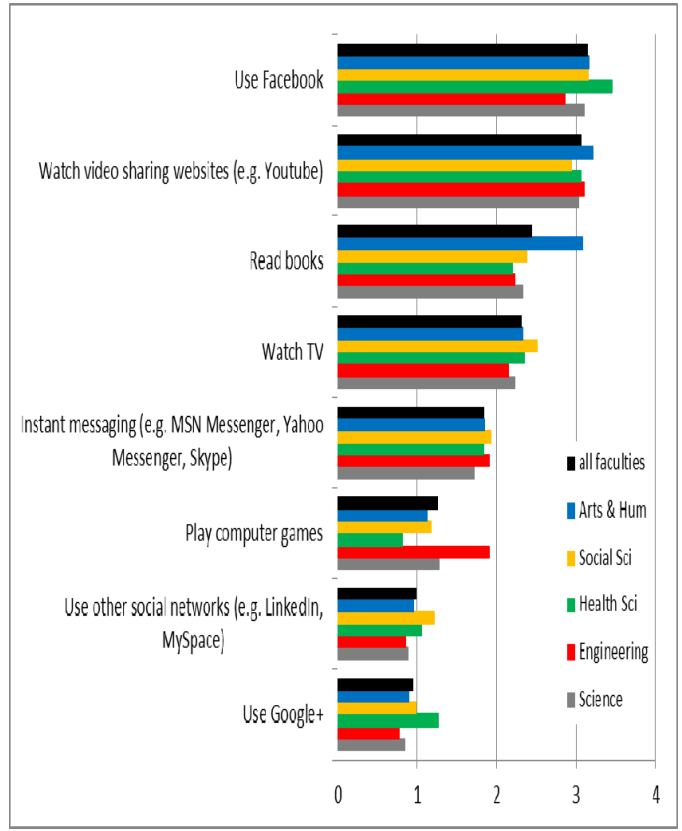

Fig. 6. Means of students' responses to the question: How often do you do the following during your free time?

IV. CONCLUSION

Looking at the survey results, it can be stated that several traditional media are still very relevant and continuing to be in high use, however, in a changing environment. Printed material and slides from the instructors as well as printed books were deemed to have high values of usage frequency and satisfaction. At the same time, new media, such as the electronic versions of material from instructors, are established and utilized with a similar intensity. It seems that these newly established media, which are based on traditional media, are very easy and comfortable to access and use and, therefore, in the future they are likely to be used more often than their traditional counterparts.